# GRAVITATIONAL-SCALAR INSTABILITY OF A COSMOLOGICAL MODEL BASED ON A TWO-COMPONENT SYSTEM OF DEGENERATE SCALARLY CHARGED FERMIONS WITH ASYMMETRIC HIGGS INTERACTION. II. WKB-APPROXIMATION.

**Yu. G. Ignat'ev**

*Solutions of the system of evolutionary equations in the short-wavelength approximation are found and studied. A connection is established between the problem of the evolution of short-wavelength gravitational-scalar perturbations and the problem of eigenvectors and eigenvalues of a linear operator. By means of an applied mathematical package, a program for calculating the amplitudes of perturbations of the cosmological model was created. It is shown that unstable short-wavelength perturbation modes arise in such a system. A numerical model for the evolution of gravitational-scalar perturbations of the cosmological model has been constructed and analyzed.*

In this part of the article, we study the complete system of linear differential equations for gravitational-scalar perturbations of a homogeneous isotropic cosmological system of degenerate scalarly charged fermions obtained in the previous part of the article in the short-wavelength WKB approximation[1]:

$$\delta\Phi'' + 2\frac{a'}{a}\delta\Phi' + [n^2 + a^2(m^2 - 3\alpha\Phi^2)]\delta\Phi + \frac{1}{2}\Phi'\mu' = -8\pi a^2 \delta\sigma^z, \qquad (I.59)$$

$$\delta\varphi'' + 2\frac{a'}{a}\delta\varphi' + [n^2 - a^2(\mathfrak{m}^2 + 3\beta\varphi^2)]\delta\varphi + \frac{1}{2}\varphi'\mu' = 8\pi a^2 \delta\sigma^\zeta, \qquad (I.60)$$

$$v = \frac{in}{8\pi a^3(\varepsilon + p)_p}\left(\delta\Phi\Phi' - \delta\varphi\varphi' + \frac{v'}{3}\right), \qquad (I.61)$$

$$8\pi a^2 \delta\varepsilon_p = \frac{a'}{a}\mu' - \Phi'\delta\Phi' + \varphi'\delta\varphi' + \frac{n^2}{3}v - a^2(m^2 - \alpha\Phi^2)\Phi\delta\Phi - a^2(\mathfrak{m}^2 - \beta\varphi^2)\varphi\delta\varphi, \qquad (I.62)$$

$$\lambda'' + 2\frac{a'}{a}\lambda' - \frac{1}{3}n^2 v = 0, \qquad (I.67)$$

$$v'' + 2\frac{a'}{a}v' + \frac{1}{3}n^2 v + 3\delta\Phi'\Phi' - 3\delta\varphi'\varphi' - 3a^2[\Phi\delta\Phi(m^2 - \alpha\Phi^2) + \varphi\delta\varphi(\mathfrak{m}^2 - \beta\varphi^2) - 8\pi\delta p_p] = 0, \qquad (I.68)$$

where $v = \lambda + \mu$.

---

[1] At the same time, we will continue the numbering of sections and formulas, and we will refer to the formula "No" of the first part of the article as (I.No).

## 4. WKB APPROXIMATION

### 4.1. WKB approximation, dimension

Let us carry out a dimensional analysis of the dynamical system under study (I.59), (I.60), (I.62), (I.67) and (I.68). It follows from (I.1) that the quantities $q^r \Phi_r$ have the dimension of mass[2]

$$[e_r \Phi] = [e_\zeta] = [m] = \left[\frac{1}{\ell}\right],$$

where $[A]$ denotes the dimension of quantity $A$, and $\ell$ here and below denotes the length scale. Next,

$$[\eta] = [x, y, z] = [\ell], \quad [a(\eta)] = [1], \quad [n] = \left[\frac{1}{\ell}\right].$$

Therefore, from the Einstein equation (45),

$$[\Lambda] = \left[\frac{1}{\ell^2}\right], \quad [\Phi, \varphi] = [1], \quad [\alpha, \beta] = \left[\frac{1}{\ell^2}\right], \quad [e_r, e_\zeta] = \left[\frac{1}{\ell^{1/2}}\right].$$

Based on Eq. (I.52), metric perturbations are dimensionless functions as well as perturbations of scalar fields. It follows from the definition of the perturbation of the Fermi momentum (I.69) that the function $\delta(\eta)$ is also dimensionless. Thus, all perturbations are dimensionless:

$$[\mu] = [\lambda] = [\nu] = [\delta\Phi] = \delta[\varphi] = [\delta] = [1].$$

We assume that the characteristic scale of the background inhomogeneity is of the order $\ell = \eta$:

$$\frac{a'}{a} \sim \frac{1}{\eta}, \quad \frac{\Phi'}{\Phi} \sim \frac{1}{\eta}, \quad \frac{\varphi'}{\varphi} \sim \frac{1}{\eta}.$$

Let us study the formulated mathematical model of the cosmological evolution of perturbations (I.59), (I.60), (I.62), (I.67), and (I.68) in the short-wave sector of perturbations (see, for example, [1]):

$$n\eta \gg 1 \Rightarrow \delta\Phi' \sim n\delta\Phi \gg \frac{a'}{a}\delta\Phi \sim \frac{1}{\eta}\delta\Phi, \quad \delta\varphi' \sim n\delta\varphi \gg \frac{a'}{a}\delta\varphi \sim \frac{1}{\eta}\delta\Phi, \tag{80}$$

etc.. In accordance with the WKB method, we represent the solutions of equations $f(\eta)$ in the form:

$$f = \tilde{f}(\eta) \cdot e^{i\int u(\eta) d\eta}, \quad (|u| \sim n \gg 1/\eta), \tag{81}$$

where $\tilde{f}(\eta)$ and $u(\eta)$ are the functions of the amplitude and eikonal of the perturbation weakly changing along with the scale factor.

---

[2] We remind you that we are using the Planck system of units, $G = c = \hbar = 1$.

### 4.2. WKB solutions of wave equations

In order not to return every time to the question of the WKB approximation, we will demonstrate the technique for solving equations in the WKB approximation using the example of finding the WKB solutions of the wave equation with respect to the scalar function $\Psi(z,\eta)$

$$\Box\Psi = 0. \tag{82}$$

In what follows, we will refer to the WKB solutions of the wave equations obtained below as basic ones. Representing $\Psi$ in the form (81), we obtain from (82) the equation:

$$\Box\Psi = \frac{1}{a^2}\left[(n^2 - u^2)\tilde{\Psi} + iu'\tilde{\Psi} + 2iu\tilde{\Psi}' + \tilde{\Psi}'' + 2\frac{a'}{a}(iu\tilde{\Psi} + \tilde{\Psi}')\right]e^{i\int u(\eta)d\eta} = 0. \tag{83}$$

Separating the orders of the WKB approximation in (83) and restricting ourselves to the first WKB approximation, we obtain the equations:

$$\begin{array}{l|ll}
\text{0th order} & \dfrac{1}{a^2}(n^2 - u^2)\tilde{\psi} & = 0, \\
\text{1st order} & \dfrac{1}{a^2}\left(u'\tilde{\psi} + 2u\tilde{\psi}' + 2\dfrac{a'}{a}u\tilde{\psi}\right) & = 0.
\end{array}$$

From the zero approximation equation we find $u = \pm n$. Substituting this solution into the equation of the first approximation, we find and, thus, find the WKB solution of the wave equation (82):

$$\Psi(z,\eta) \simeq \frac{\Psi_n^0}{a}e^{in(z\pm\eta)}, \tag{84}$$

which describes a retarded and advanced wave propagating at the speed of light, the amplitude of which falls inversely proportional to the scale factor. Consider now the wave equation for a massive vacuum scalar field:

$$\Box\Psi + m^2(\eta)\Psi = 0, \tag{85}$$

where the massive term is also a slowly varying quantity $m' \sim m/\eta$. Proceeding similarly to the previous one, instead of (83) we obtain the following equation:

$$\Box\Psi + m^2\Phi = \frac{1}{a^2}\left[(n^2 - u^2 + a^2m^2)\tilde{\Psi} + iu'\tilde{\Psi} + 2iu\tilde{\Psi}' + \tilde{\Psi}'' + 2\frac{a'}{a}(iu\tilde{\Psi} + \tilde{\Psi}')\right]e^{i\int u(\eta)d\eta} = 0. \tag{86}$$

Considering the fact that the value $a(\eta)m(\eta)$ can become on the order of and even larger with increasing time, we must keep this term in the zero-order equation. Thus, we have equations in the first WKB approximation:

$$\begin{array}{l|ll}
\text{0th order} & \dfrac{1}{a^2}(n^2 - u^2 + a^2m^2)\tilde{\psi} & = 0, \\
\text{1st order} & \dfrac{1}{a^2}\left(u'\tilde{\psi} + 2u\tilde{\psi}' + 2\dfrac{a'}{a}u\tilde{\psi}\right) & = 0,
\end{array}$$

whence for the zero WKB approximation we find:

$$u = \pm\sqrt{n^2 + a^2 m^2}. \qquad (87)$$

Substituting solution (87) into the equation of the first WKB approximation, we easily find:

$$\Psi(z,\eta) \simeq \frac{\Psi_n^0}{a(n^2 + a^2 m^2)^{1/4}} e^{inz \pm i\int\sqrt{n^2 + a^2 m^2}\,d\eta}. \qquad (88)$$

At $am/n \to 0$, solution (88) goes over to solution (84). Since the integrand in the exponent (88) serves as the frequency of the wave in the obtained solution, we define the phase and group velocity of the wave in the standard way[3]:

$$v_f = \frac{\omega(n)}{n} = \sqrt{1 + a^2 m^2 / n^2} > 1, \quad v_g = \frac{\partial \omega(n)}{\partial n} = \frac{1}{\sqrt{1 + a^2 m^2 / n^2}} < 1. \qquad (89)$$

### 4.3. WKB solutions for constraint equations

Note that, in contrast to the variables $\delta\Phi(\eta)$, $\delta\varphi(\eta)$, $\lambda(\eta)$, and $\nu(\eta)$, the variable $\delta(\eta)$ enters the dynamic equations only algebraically, which makes it possible to reduce the number of dynamic functions to four. To find $\delta(\eta)$, one can use the Einstein equation (I.62), in the left side of which it is necessary to substitute the expression $\delta\varepsilon_p$ from (I.78) and, according to the WKB approximation, discard small quantities of the order of $n/\ell_0$. As a result, under conditions $\varepsilon_p \neq 0$, $\delta\varepsilon_p \neq 0$ (I.65), we obtain an algebraic connection:

$$\delta = \frac{\frac{n^2}{3}\nu - a^2[(\varepsilon_p^\Phi + m^2\Phi - \alpha\Phi^3)\delta\Phi + (\varepsilon_p^\varphi + \mathfrak{m}^2\varphi - \beta\varphi^3)\delta\varphi]}{8\pi a^2 \varepsilon_p^\delta} \equiv \frac{n^2}{24\pi a^2 \varepsilon_p^\delta}\nu - \Delta_\Phi \delta\Phi - \Delta_\varphi \delta\varphi, \qquad (90)$$

where the following notation are introduced:

$$\varepsilon_p^\delta = \frac{1}{\pi^2}(e_z^4 \Phi^4 \psi_z^3 \sqrt{1+\psi_z^2} + e_\zeta^4 \varphi^4 \psi_\zeta^3 \sqrt{1+\psi_\zeta^2}) > 0,$$

$$\varepsilon_p^\Phi = \frac{e_z^4 \Phi^3}{2\pi^2} F_1(\psi_z), \quad \varepsilon_p^\varphi = \frac{e_\zeta^4 \varphi^3}{2\pi^2} F_1(\psi_\zeta), \quad \Delta_\Phi = \frac{\varepsilon_p^\Phi}{8\pi \varepsilon_p^\delta}, \quad \Delta_\varphi = \frac{\varepsilon_p^\varphi}{8\pi \varepsilon_p^\delta}. \qquad (91)$$

Note that all coefficients in a linear relationship $\delta(\nu, \delta\Phi, \delta\varphi)$ are dimensionless quantities, while the quantities $\varepsilon_p^\delta$, $\varepsilon_p^\Phi$ and $\varepsilon_p^\varphi$ have the dimension $\ell^{-2}$. Next, we must substitute the obtained expression for $\delta$ (90) into the expressions for the perturbations of scalar charge densities $s^z$ (I.76) and $s^\zeta$ (I.77), as well as into the expression for the perturbation of the pressure of the statistical system (I.79). The expressions thus obtained must, in turn, be substituted into the corresponding

---

[3] see, for example, [3].

equations for perturbations of scalar fields (I.59) and (I.60) and Einstein's equation for perturbations (I.68). Thus, we get, for example:

$$\delta\sigma^z = \frac{e_z^4 \Phi^3 \psi_z^2}{48\pi^3 a^2 \varepsilon_p^\delta \sqrt{1+\psi_z^2}} n^2 \nu + S_\Phi^z \delta\Phi + S_\varphi^z \delta\varphi,$$

(92)

$$S_\Phi^z = \frac{e_z^4 \Phi^2}{2\pi^2}\left(3F_1(\psi_z) - \frac{\psi_z^3}{\sqrt{1+\psi_z^2}} - \frac{\psi_z^2}{\sqrt{1+\psi_z^2}}\Delta_\Phi\right), \quad S_\varphi^z = -\frac{e_z^4 \Phi^2 \psi_z^2}{2\pi^2 \sqrt{1+\psi_z^2}}\Delta_\varphi,$$

$$\delta\sigma^\zeta = \frac{e_\zeta^4 \varphi^3 \psi_\zeta^2}{48\pi^3 a^2 \varepsilon_p^\delta \sqrt{1+\psi_\zeta^2}} n^2 \nu + S_\Phi^\zeta \delta\Phi + S_\varphi^\zeta \delta\varphi,$$

(93)

$$S_\varphi^\zeta = \frac{e_\zeta^4 \varphi^2}{2\pi^2}\left(3F_1(\psi_\zeta) - \frac{\psi_\zeta^3}{\sqrt{1+\psi_\zeta^2}} - \frac{\psi_\zeta^2}{\sqrt{1+\psi_\zeta^2}}\Delta_\varphi\right), \quad S_\Phi^\zeta = -\frac{e_\zeta^4 \varphi^2 \psi_\zeta^2}{2\pi^2 \sqrt{1+\psi_\zeta^2}}\Delta_\Phi,$$

$$\delta p_p = \frac{p_p^\delta n^2}{24\pi^3 a^2 \varepsilon_p^\delta}\nu + P^\Phi \delta\Phi + P^\varphi \delta\varphi, \quad p_p^\delta = \frac{1}{\pi^2}\left(\frac{e_z^4 \Phi^4 \psi_z^4}{\sqrt{1+\psi_z^2}} + \frac{e_\zeta^4 \varphi^4 \psi_\zeta^4}{\sqrt{1+\psi_\zeta^2}}\right) > 0,$$

(94)

$$P^\Phi = \frac{e_z^4 \Phi^3}{2\pi^2} F_1(\psi_z) - p_p^\delta \Delta_\Phi, \quad P^\varphi = \frac{e_\zeta^4 \varphi^3}{2\pi^2} F_1(\psi_\zeta) - p_p^\delta \Delta_\varphi.$$

All coefficients in linear relations $\delta\sigma^{z,\zeta}(\nu,\delta\Phi,\delta\varphi)$ and $\delta p_p(\nu,\delta\Phi,\delta\varphi)$ have the dimension $\ell^{-2}$, while $p_p^\delta$ is a dimensionless value.

**4.4. Asymptotic values in Eqns (91)–(94)**

In what follows, we will need the asymptotic values of functions (90) – (94) in the nonrelativistic ($\psi_r \to 0$) and ultrarelativistic ($\psi_r \to \infty$) limits. Note that these limits are also equivalent to large and small, respectively, values of charges, or small and large, respectively, values of Fermi momenta. Using the corresponding asymptotics for the functions (I.22) and (I.23), we obtain the following asymptotics.

Nonrelativistic limit $\psi_r \to 0 \leftrightarrow$ heavy charges $|e_r| \gg \pi_r$

$$\varepsilon_p^\delta \simeq \frac{2(|e_z\Phi|\pi_z^3 + |e_\zeta\varphi|\pi_\zeta^3)}{\pi^2}, \quad p_p^\delta \simeq \frac{2}{\pi^2}(\pi_z^4 + \pi_\zeta^4), \quad \varepsilon_p^\Phi \simeq \frac{|e_z|}{3\pi^2}\mathrm{sgn}(\Phi)\pi_z^3, \quad \varepsilon_p^\varphi \simeq \frac{|e_\zeta|}{3\pi^2}\mathrm{sgn}(\varphi)\pi_\zeta^3,$$

$$\Delta_\Phi \simeq \frac{|e_z|\mathrm{sgn}(\Phi)\pi_z^3}{24\pi(\pi_z^3|e_z\Phi|+\pi_\zeta^3|e_\zeta\varphi|)}, \quad \Delta_\varphi \simeq \frac{|e_\zeta|\mathrm{sgn}(\varphi)\pi_\zeta^3}{24\pi(\pi_z^3|e_z\Phi|+\pi_\zeta^3|e_\zeta\varphi|)}, \quad P^\Phi \simeq \frac{|e_z|\pi_z^3}{3\pi^2}\mathrm{sgn}(\Phi),$$

$$S^z_\Phi \simeq -\frac{|e_z|^3 \pi_z^5 \mathrm{sgn}(\Phi)}{596\pi^3(\pi_z^3|e_z\Phi|+\pi_\zeta^3|e_\zeta\varphi|)}, \quad S^\zeta_\varphi \simeq -\frac{|e_\zeta|^3 \pi_\zeta^5}{596\pi^3} \frac{\mathrm{sgn}(\varphi)}{\pi_z^3|e_z\Phi|+\pi_\zeta^3|e_\zeta\varphi|}, \quad (95)$$

$$S^\zeta_\Phi \simeq -\frac{e_\zeta^2 |e_z| \pi_\zeta^2 \pi_z^3}{4\pi^2}\mathrm{sgn}(\Phi), \quad S^z_\varphi \simeq -\frac{e_z^2 |e_\zeta| \pi_z^2 \pi_\zeta^3}{4\pi^2}\mathrm{sgn}(\varphi), \quad P^\varphi \simeq \frac{|e_\zeta|\pi_\zeta^3}{3\pi^2}\mathrm{sgn}(\varphi).$$

Ultrarelativistic limit $\psi_r \to \infty \leftrightarrow$ small charges $|\mathbf{e_r}| \ll \pi_\mathbf{r}$

$$\varepsilon^\delta_p \simeq \frac{1}{\pi^2}(\pi_z^4+\pi_\zeta^4), \quad p^\delta_p \simeq \frac{1}{\pi^2}(|e_z\Phi|\pi_z^3+|e_\zeta\varphi|\pi_\zeta^3), \quad \varepsilon^\Phi_p \simeq \frac{e_z^2\Phi}{2\pi^2}\pi_z^2, \quad \varepsilon^\varphi_p \simeq \frac{e_\zeta^2\varphi}{2\pi^2}\pi_\zeta^2,$$

$$\Delta_\Phi \simeq \frac{e_z^2\Phi\pi_z^2}{2(\pi_z^4+\pi_\zeta^4)}, \quad \Delta_\varphi \simeq \frac{e_\zeta^2\varphi\pi_\zeta^2}{2(\pi_z^4+\pi_\zeta^4)}, \quad S^z_\Phi \simeq \frac{|e_z^3|\Phi\pi_z}{4\pi^2(\pi_z^4+\pi_\zeta^4)}, \quad S^\zeta_\varphi \simeq \frac{|e_\zeta^3|\varphi\pi_\zeta}{4\pi^2(\pi_z^4+\pi_\zeta^4)},$$

$$S^\zeta_\Phi \simeq -\frac{|e_\zeta^3\varphi| e_z^2\Phi\pi_\zeta\pi_z^2}{4\pi^2(\pi_z^4+\pi_\zeta^4)}, \quad S^z_\varphi \simeq -\frac{|e_z^3\Phi|e_\zeta^2\varphi\pi_z\pi_\zeta^2}{4\pi^2(\pi_z^4+\pi_\zeta^4)}, \quad P^\Phi \simeq \frac{e_z^2\Phi\pi_z^3}{2\pi^2}, \quad P^\varphi \simeq \frac{e_\zeta^2\varphi\pi_\zeta^3}{2\pi^2}. \quad (96)$$

### 4.5. Perturbation equations in WKB terms

Representing, for example, $\delta\Phi(\eta)$ in the form (81), we find:

$$\delta\Phi(\eta) = \delta\tilde\Phi(\eta)\cdot e^{i\int u(\eta)d\eta}, \quad \delta\Phi' = [\delta\tilde\Phi' + iu\delta\tilde\Phi]e^{i\int u(\eta)d\eta},$$

$$\delta\Phi'' = [\delta\tilde\Phi'' + 2iu\delta\tilde\Phi' + iu'\delta\tilde\Phi - u^2\delta\tilde\Phi]e^{i\int u(\eta)d\eta}. \quad (97)$$

Substituting the obtained expressions for perturbations of macroscopic scalars (92) - (94) into the equations for perturbations of scalar (I.59), (I.60) and gravitational fields (67) and then substituting the perturbations $\delta\Phi$, $\delta\varphi$, $\nu$ and $\lambda$ in the form (81) into the system (I.59), (I.60), (I.67) and (I.68), taking into account relations (97), we obtain a system of ordinary differential equations for the amplitudes $\delta\tilde\Phi$, $\delta\tilde\varphi$, $\tilde\nu$, $\tilde\lambda$ and the eikonal function $u(\eta)$.
:

$$[n^2 - u^2 + a^2(m^2 - 3\alpha\Phi^2 + 8\pi S^z_\Phi)]\delta\tilde\Phi + \frac{e_z^4\Phi^3\psi_z^2}{6\pi^2\varepsilon^\delta_p\sqrt{1+\psi_z^2}}n^2\tilde\nu - 8\pi a^2 S^z_\varphi\delta\tilde\varphi$$

$$+i\left[2u\delta\tilde\Phi' - u'\delta\tilde\Phi + 2\frac{a'}{a}u\delta\tilde\Phi - \frac{1}{2}\Phi'u(\tilde\lambda-\tilde\nu)\right] + \left[\delta\tilde\Phi'' + 2\frac{a'}{a}\delta\tilde\Phi' - \frac{1}{2}\Phi'\tilde\mu'\right] = 0, \quad (98)$$

$$[n^2 - u^2 - a^2(\mathfrak{m}^2 - 3\beta\varphi^2 + 8\pi S^\zeta_\varphi)]\delta\tilde\varphi - \frac{e_\zeta^4\varphi^3\psi_\zeta^2}{6\pi^2\varepsilon^\delta_p\sqrt{1+\psi_\zeta^2}}n^2\tilde\nu + 8\pi a^2 S^\zeta_\Phi\delta\tilde\Phi$$

$$+i\left[2u\delta\tilde{\varphi}' - u'\delta\tilde{\varphi} + 2\frac{a'}{a}u\delta\tilde{\varphi} - \frac{1}{2}\varphi'u(\tilde{\lambda}-\tilde{v})\right] + \left[\delta\tilde{\varphi}'' + 2\frac{a'}{a}\delta\tilde{\varphi}' - \frac{1}{2}\varphi'\tilde{\mu}'\right] = 0, \qquad (99)$$

$$-u^2\tilde{\lambda} - \frac{1}{3}n^2\tilde{v} + i\left[2u\tilde{\lambda}' - u'\tilde{\lambda} + 2\frac{a'}{a}u\tilde{\lambda}\right] + \left[\tilde{\lambda}'' + 2\frac{a'}{a}\tilde{\lambda}'\right] = 0, \qquad (100)$$

$$\left[n^2\left(\frac{1}{3} + \frac{p_p^\delta}{\varepsilon_p^\delta}\right) - u^2\right]\tilde{v} - 3a^2[\Phi(\mathfrak{m}^2 - \alpha\Phi^2) - 8\pi P^\Phi]\delta\tilde{\Phi} - 3a^2[\varphi(\mathfrak{m}^2 - \beta\varphi^2) - 8\pi P^\varphi]\delta\tilde{\varphi}$$

$$+i\left[2u\tilde{v}' - u'\tilde{v} + 2\frac{a'}{a}u\tilde{v} + 3u\delta\tilde{\Phi}\Phi' - 3u\delta\tilde{\varphi}\varphi\right] + \left[\tilde{\mu}'' + 2\frac{a'}{a}\tilde{\mu}' + 3\Phi'\delta\tilde{\Phi}' - 3\varphi'\delta\tilde{\varphi}'\right] = 0. \qquad (101)$$

## 5. SOLUTION OF EQUATIONS FOR PERTURBATIONS BY THE WKB METHOD

### 5.1. Dispersion equation: zero order WKB

Equations (98) - (101) in the zero order WKB represent a system of homogeneous algebraic equations for the amplitudes of perturbations:

$$\mathbf{A} \cdot \left[\tilde{\Phi}, \tilde{\varphi}, \tilde{\lambda}, \tilde{v}\right]^\mathrm{T} = 0, \qquad (102)$$

where T is the transposition sign and $\mathbf{A}$ is a square matrix of dimension 4.

$$A = \begin{bmatrix} n^2 - u^2 + \\ +a^2(\mathfrak{m}^2 - 3\alpha\Phi^2 + 8\pi S_\Phi^z) & -8\pi a^2 S_\varphi^z & 0 & \dfrac{e_z^4 \Phi^3 \psi_z^2}{6\pi^2 \varepsilon_p^\delta \sqrt{1+\psi_z^2}} n^2 \\ \\ 8\pi a^2 S_\Phi^\zeta & \begin{array}{c} n^2 - u^2 - \\ -a^2(\mathfrak{m}^2 - 3\beta\varphi^2 + 8\pi S_\varphi^\zeta) \end{array} & 0 & -\dfrac{e_\zeta^4 \varphi^3 \psi_\zeta^2}{6\pi^2 \varepsilon_p^\delta \sqrt{1+\psi_\zeta^2}} n^2 \\ \\ 0 & 0 & -u^2 & -\dfrac{1}{3}n^2 \\ \\ \begin{array}{c} -3a^2[\Phi(\mathfrak{m}^2 - \alpha\Phi^2) - \\ -8\pi P^\Phi] \end{array} & \begin{array}{c} -3a^2[\varphi(\mathfrak{m}^2 - \beta\varphi^2) - \\ -8\pi P^\varphi] \end{array} & 0 & \left(\dfrac{1}{3} + \dfrac{p_p^\delta}{\varepsilon_p^\delta}\right)n^2 - u^2 \end{bmatrix}. \qquad (103)$$

A necessary and sufficient condition for the nontrivial solvability of system (103) is the equality to zero of the determinant of the matrix $\mathbf{A}$:

$$\det(\mathbf{A}(u,n)) = 0. \qquad (104)$$

Equations of type (105) that establish relationships of the form $u = u(n,t)$, i.e., relationships between the wave vector and the oscillation frequency, are called dispersion equations in plasma theory (see, for example, [2]). The dispersion equation (105) is solved by eight symmetric eikonal functions $\pm u_{(k)}(n,\eta), (k = \overline{1..4})$, whose successive substitution into equations (103) will give us the fundamental solution $(\tilde{\Phi}_i(n), \tilde{\varphi}_i(n), \tilde{\lambda}_i(n), \tilde{n}u_i(n))$. The resulting algebraic solutions then need to be substituted into the first-order differential equations of the WKB approximation to determine the dependence of the solutions on the time variable.

The type (104) equations, which set constraints $u = u(n,t)$, i.e., those between the wave vector and vibration frequency, are called dispersion equations in the plasma theory (see, for example, [2]). The solution of Eq. (104), is eight symmetric eikonal functions $\pm u_{(k)}(n,\eta), (k = \overline{1..4})$ [4]. Their successive insertion in Eq. (102) yields the fundamental solution of $(\tilde{\Phi}_i(n), \tilde{\varphi}_i(n), \tilde{\lambda}_i(n), \tilde{n}u_i(n))$. Algebraic solutions should be then inserted in differential equations of the first-order WKB approximation in order to determine the time-dependent solutions.

Since the third column of the matrix $\mathbf{A}$ contains only one nonzero element $a_{31} = -u^2$, then, firstly, the dispersion equation (105) can be represented as:

$$u^2 \det(\overline{\mathbf{A}}) = 0, \qquad (105)$$

where $\overline{\mathbf{A}}$ is the is the square matrix of dimension 3,

$$\overline{A} = \begin{bmatrix} n^2 - u^2 + \\ +a^2(m^2 - 3\alpha\Phi^2 + 8\pi S_\Phi^z) & -8\pi a^2 S_\varphi^z & \dfrac{e_z^4 \Phi^3 \psi_z^2}{6\pi^2 \varepsilon_p^\delta \sqrt{1+\psi_z^2}} n^2 \\ \\ 8\pi a^2 S_\Phi^\zeta & \begin{matrix} n^2 - u^2 - \\ -a^2(\mathfrak{m}^2 - 3\beta\varphi^2 + 8\pi S_\varphi^\zeta) \end{matrix} & -\dfrac{e_\zeta^4 \varphi^3 \psi_\zeta^2}{6\pi^2 \varepsilon_p^\delta \sqrt{1+\psi_\zeta^2}} n^2 \\ \\ -3a^2[\Phi(m^2 - \alpha\Phi^2) - 8\pi P^\Phi] & -3a^2[\varphi(\mathfrak{m}^2 - \beta\varphi^2) - 8\pi P^\varphi] & \left(\dfrac{1}{3} + \dfrac{p_p^\delta}{\varepsilon_p^\delta}\right)n^2 - u^2 \end{bmatrix}. \qquad (106)$$

Second, from equation (106) we immediately find the trivial solution:

$$u_{(3)}^\pm = 0. \qquad (107)$$

According to (103) and (104), this solution in the zero WKB approximation corresponds to an arbitrary function $\tilde{\lambda}(\eta)$:

$$(107) \Rightarrow \forall \tilde{\lambda} \in \mathbb{R}, \quad \delta\tilde{\Phi} = \delta\tilde{\varphi} = \tilde{v} = 0. \qquad (108)$$

The same zero mode also appears in Lifshitz's theory of gravitational instability [1], [3], if this theory is reformulated in WKB terms. As shown in [2], this mode is excluded by admissible coordinate transformations. In what follows, we will omit the zero mode. The dispersion equation for the rest of the perturbation modes ($\delta\tilde{\Phi}_i(n)$, $\delta\tilde{\varphi}_i(n)$ and $\tilde{v}_i(n)$) has the form relative $u^2$ to

$$\det(\bar{\mathbf{A}}) = 0, \qquad (109)$$

where

$$\bar{A} = \begin{bmatrix} n^2 - u^2 + \gamma_{11} & \gamma_{12} & \gamma_{13}n^2 \\ \gamma_{21} & n^2 - u^2 + \gamma_{22} & \gamma_{23}n^2 \\ \gamma_{31} & \gamma_{32} & \gamma_{33}n^2 - u^2 \end{bmatrix}, \qquad (110)$$

and the following notation are introduced:

$$\gamma_{11} \equiv +a^2(\mathfrak{m}^2 - 3\alpha\Phi^2 + 8\pi S_\Phi^z), \quad \gamma_{22} \equiv -a^2(\mathfrak{m}^2 - 3\beta\varphi^2 + 8\pi S_\varphi^\zeta),$$

$$\gamma_{33} \equiv \frac{1}{3} + \frac{p_p^\delta}{\varepsilon_p^\delta}, \quad \gamma_{12} \equiv -8\pi a^2 S_\varphi^z, \quad \gamma_{13} \equiv \frac{e_z^4 \Phi^3 \psi_z^2}{6\pi^2 \varepsilon_p^\delta \sqrt{1+\psi_z^2}},$$

$$\gamma_{21} \equiv 8\pi a^2 S_\Phi^\zeta, \quad \gamma_{31} \equiv -3a^2[\Phi(\mathfrak{m}^2 - \alpha\Phi^2) - 8\pi P^\Phi],$$

$$\gamma_{32} \equiv -3a^2[\varphi(\mathfrak{m}^2 - \beta\varphi^2) - 8\pi P^\varphi], \quad \gamma_{23} \equiv -\frac{e_\zeta^4 \varphi^3 \psi_\zeta^2}{6\pi^2 \varepsilon_p^\delta \sqrt{1+\psi_\zeta^2}}. \qquad (111)$$

---

[4] different signs $u$ correspond to retarded and advanced waves.

Thus, in the zero WKB approximation, equations (103) with respect to variables $\tilde{\Phi}, \tilde{\varphi}, \tilde{v}$ can be written in the form

$$\bar{\mathbf{A}} \cdot [\tilde{\Phi}, \tilde{\varphi}, \tilde{v}]^T = 0. \qquad (112)$$

Calculating the matrix determinant (110) and introducing the variable

$$x \equiv \frac{u^2}{n^2} \sim 1,$$

we obtain the dispersion equation for the remaining vibration modes in the form of a polynomial in even degree $n$:

$$\det(\bar{\mathbf{A}}) = (x-1)^2(\gamma_{33}-x)n^6 + (\gamma_{11}+\gamma_{22})(x-1)(\gamma_{33}-x)n^4 + [-(\gamma_{11}\gamma_{22}-\gamma_{12}\gamma_{21})(x-\gamma_{33})$$

$$+\gamma_{13}\gamma_{31}+\gamma_{23}\gamma_{32}]n^2 - \gamma_{11}\gamma_{23}\gamma_{32}+\gamma_{13}\gamma_{21}\gamma_{32}+\gamma_{12}\gamma_{23}\gamma_{31}-\gamma_{13}\gamma_{22}\gamma_{31} = 0. \qquad (113)$$

In the general case, the solution of the dispersion equation (112) is relative to the rest of the perturbation modes ($\delta\tilde{\Phi}_i(n)$, $\delta\tilde{\varphi}_i(n)$ and $\tilde{v}_i(n)$), of course, formally can be found using the Cardan formula, but this result will be extremely cumbersome and not suitable for research.

## 5.2. Numerical solution of Eq. (112)

In earlier works of the Author, the dispersion equation was solved analytically for small values of scalar charges:

$$n^2 \gg e_{(a)}^4, \quad n^2 \gtrsim a^2\{m^2\Phi, \alpha\Phi^3, \mathfrak{m}^2\varphi, \beta\varphi^3\}, \quad a^2\{m^2, \mathfrak{m}^2\} \gg e_{(a)}^4. \tag{114}$$

In this case, the dispersion equation (113) reduces to the product of equations of the 1st and 2nd orders, which are easily solved analytically. However, conditions (114) do not allow a sufficiently complete analysis of the problem. In earlier works of the Author, the dispersion equation (109) was solved by numerical methods in the applied mathematical package Maple. As it turned out, the solution of this equation, taking into account the fact that its coefficients (111), in turn, are determined by numerical background equations (I.47) - (I.51), requires rather large time resources in Maple. Therefore, in this work, we will find numerical solutions by a different method.

It is easy to see that equations (112) can be written in the form of equations for the eigenvectors of the matrix

$$\mathbf{A}_0 \mathbf{X} = k\mathbf{X}; \quad (k = u^2), \tag{115}$$

where $\mathbf{X} = (\tilde{\Phi}, \tilde{\varphi}, \tilde{\nu})^T$ and

$$A_0 = \begin{bmatrix} n^2 + \gamma_{11} & \gamma_{12} & \gamma_{13}n^2 \\ \gamma_{21} & n^2 + \gamma_{22} & \gamma_{23}n^2 \\ \gamma_{31} & \gamma_{32} & \gamma_{33}n^2 \end{bmatrix}. \tag{116}$$

Therefore, the problem of finding solutions to equations for perturbations in the zeroth WKB order is reduced to the problem of eigenvectors of matrix $\mathbf{A}_0$ (116), and the eigenvalues of this matrix determine the eigen-frequencies $\omega(t)$ of oscillations and increments/decrements of eigen-oscillations according to the formulas (pay attention to the sign of the increment):

$$\omega^{\pm}{}_\alpha(t) = \pm e^{\xi(t)} \mathfrak{R}(\sqrt{k_\alpha(t)}) \equiv \pm \omega_\alpha(t), \quad \gamma^{\pm}{}_\alpha(t) = \mp e^{\xi(t)} \mathfrak{I}(\sqrt{k_\alpha(t)}) \equiv \mp \gamma_\alpha(t), \quad (\alpha = \overline{1,3}), \tag{117}$$

and wave amplitudes (Eq. (81)) are derived from

$$\mathbf{X}_\alpha(t) = \sum_{\pm} \tilde{\mathbf{X}}_\alpha^{\pm}(t) \exp\left(\pm \int \gamma_\alpha(t)dt\right) e^{\pm i \int \omega_\alpha(t)dt}, \quad \left(n \int_{t_0}^{t} e^{\xi(t)} dt \gg 1\right), \tag{118}$$

where we transfer from the time variable $\eta$ to cosmological time $t$:

$$\eta = \int_{t_0}^{t} \frac{dt}{a(t)} \equiv \int_{t_0}^{t} e^{-\xi(t)} dt. \tag{119}$$

As a result, the perturbations $\mathbf{F}_n(z,t) = [\delta\Phi, \delta\varphi, \delta\nu]$ are described by the following expression

$$\mathbf{F}_n(z,t) = \sum_{\pm} \tilde{\mathbf{X}}_\alpha^{\pm}(t) \exp\left(\pm \int \gamma_\alpha(t)dt\right) e^{inz \pm \int \omega_\alpha(t)dt}. \tag{120}$$

It turns out that the latest versions of the Maple package contain efficient software tools for numerically solving the problem of eigenvectors and matrix eigenvalues. At the same time, the speed of the numerical solution of this problem, in fact, is an order of magnitude higher than the speed of calculating the roots of the cubic equation. In this case, it is only necessary to correct minor inaccuracies in the decision inference program. We will use these means to determine the frequencies, increments, and amplitudes of perturbations.

**5.3. First and second orders**

The dependences of the amplitudes of the found modes on the time variable are determined by the solutions of first-order differential equations in the WKB approximation. In the first order WKB, from (98) - (101) we have the following system of ordinary differential equations of the first order:

$$2u\delta\tilde{\Phi}' - u'\delta\tilde{\Phi} + 2\frac{a'}{a}u\delta\tilde{\Phi} - \frac{1}{2}\Phi'u(\tilde{\lambda} - \tilde{\nu}) = 0, \tag{121}$$

$$2u\delta\tilde{\varphi}' - u'\delta\tilde{\varphi} + 2\frac{a'}{a}u\delta\tilde{\varphi} - \frac{1}{2}\varphi'u(\tilde{\lambda} - \tilde{\nu})] = 0, \tag{122}$$

$$2u\tilde{\lambda}' - u'\tilde{\lambda} + 2\frac{a'}{a}u\tilde{\lambda} = 0, \tag{123}$$

$$2u\tilde{\nu}' - u'\tilde{\nu} + 2\frac{a'}{a}u\tilde{\nu} + 3u\delta\tilde{\Phi}\Phi' - 3u\delta\tilde{\varphi}\varphi = 0. \tag{124}$$

We note, firstly, that the solution of the dispersion equation $u^{\pm}_{(3)} = 0$ turns these equations into identities, and secondly, that equation (124) is easily integrated, since it contains a total differential:

$$u^2 a^2 \tilde{\lambda} = C_1. \tag{125}$$

Due to this circumstance, in a number of cases it is necessary to retain terms of the second order with respect to the WKB parameter. Thus, we get the equations:

$$\delta\tilde{\Phi}'' + 2\frac{a'}{a}\delta\tilde{\Phi}' - \frac{1}{2}\Phi'\tilde{\mu}' = 0, \tag{126}$$

$$\delta\tilde{\varphi}'' + 2\frac{a'}{a}\delta\tilde{\varphi}' - \frac{1}{2}\varphi'\tilde{\mu}' = 0, \tag{127}$$

$$\tilde{\lambda}'' + 2\frac{a'}{a}\tilde{\lambda}' = 0, \tag{128}$$

$$\tilde{\mu}'' + 2\frac{a'}{a}\tilde{\mu}' + 3\Phi'\delta\tilde{\Phi}' - 3\varphi'\delta\tilde{\varphi}' = 0. \tag{129}$$

These equations do not include the eikonal function, so we can immediately obtain in the case $u = 0$ instead of (125) the solution:

$$\tilde{\lambda} = \frac{C_1}{a^2}. \qquad (130)$$

Since at $u \not= 0$ we have obtained an explicit solution for (130), we can simplify the system of equations (121) - (124) by subtracting equation (124) from equation (123) and then returning to the variable :

$$2u\delta\tilde{\Phi}' - u'\delta\tilde{\Phi} + 2\frac{a'}{a}u\delta\tilde{\Phi} - \frac{1}{2}\Phi'u\tilde{\mu} = 0, \qquad (131)$$

$$2u\delta\tilde{\varphi}' - u'\delta\tilde{\varphi} + 2\frac{a'}{a}u\delta\tilde{\varphi} - \frac{1}{2}\varphi'u\tilde{\mu} = 0, \qquad (132)$$

$$2u\tilde{\mu}' - u'\tilde{\mu} + 2\frac{a'}{a}u\tilde{\mu} - 3u\delta\tilde{\Phi}\Phi' + 3u\delta\tilde{\varphi}\varphi = 0. \qquad (133)$$

Equations (131)–(133) can be written in a more reduced from

$$\frac{d}{d\eta}\frac{a\delta\tilde{\Phi}}{\sqrt{u}} = \frac{a\Phi'}{4\sqrt{u}}\tilde{\mu}; \quad \frac{d}{d\eta}\frac{a\delta\tilde{\varphi}}{\sqrt{u}} = \frac{a\varphi'}{4\sqrt{u}}\tilde{\mu}; \quad \frac{d}{d\eta}\frac{a\tilde{\mu}}{\sqrt{u}} = -\frac{3a}{2\sqrt{u}}(\Phi'\delta\tilde{\Phi}' - \varphi'\delta\tilde{\varphi}'). \qquad (134)$$

Further simplification of equations (134) to the amplitudes of perturbations $\tilde{\mu} = \tilde{\lambda} - \tilde{\nu}$ is not possible; their analytical study due, firstly, to the absence of small or large parameters in them and, secondly, the absence of analytical background solutions for the functions $a(\eta), \Phi(\eta), \varphi(\eta)$ is also ineffective. These functions that vary slightly with time can only be found by numerical methods. But, in principle, this is not particularly necessary, since the main information about the behavior of perturbations $\mu(\eta, z)$, $\delta\Phi(\eta, z)$, $\delta\varphi(\eta, z)$ is carried by six eikonal functions $u^{\pm}_{(\alpha)}$, $(\alpha = \overline{1,3})$. We only note that in the case of constant values of the unperturbed potentials of scalar fields $\Phi, \varphi$, which are just achieved at stable singular points of the asymmetric vacuum doublet [4], Eqs. (135) are easily integrated:

$$\delta\tilde{\Phi} = \frac{C_2\sqrt{u}}{a}, \quad \delta\tilde{\varphi} = \frac{C_3\sqrt{u}}{a}, \quad \tilde{\mu} = \frac{C_4\sqrt{u}}{a}, \quad (\Phi = \Phi_0, \varphi = \varphi_0). \qquad (135)$$

## 6. NUMERICAL MODELING

### 6.1. Notes to numerical modeling

Before moving on to numerical modeling, let us make a few remarks.

1. Macroscopic scalars (90) - (94), which determine the coefficients $\gamma_{\alpha\beta}$ of the matrix $\mathbf{A}_0$ of equations (112) with respect to eigenvectors and eigenvalues, in turn, are determined by the solutions of the system of background equations (I.47)–(I.51)

$$\dot{\xi} = H, \quad \dot{\Phi} = Z, \quad \dot{\varphi} = z, \qquad (I.47)$$

$$\dot{H} = -\frac{Z^2}{2} + \frac{z^2}{2} - \frac{4}{3\pi}e_z^4\Phi^4\psi_z^3\sqrt{1+\psi_z^2} - \frac{4}{3\pi}e_\zeta^4\varphi^4\psi_\zeta^3\sqrt{1+\psi_\zeta^2}, \qquad (I.48)$$

$$\dot{Z} = -3HZ - m^2\Phi + \Phi^3\left(\alpha - \frac{4e_z^4}{\pi}F_1(\psi_z)\right), \tag{I.49}$$

$$\dot{z} = -3Hz + \mathfrak{m}^2\varphi - \varphi^3\left(\beta - \frac{4e_\zeta^4}{\pi}F_1(\psi_\zeta)\right), \tag{I.50}$$

$$3H^2 - \Lambda - \frac{Z^2}{2} + \frac{z^2}{2} - \frac{m^2\Phi^2}{2} + \frac{\alpha\Phi^4}{4} - \frac{\mathfrak{m}^2\varphi^2}{2} + \frac{\beta\varphi^4}{4} - \frac{e_z^4\Phi^4}{\pi}F_2(\psi_z) - \frac{e_\zeta^4\varphi^4}{\pi}F_2(\psi_\zeta) = 0. \tag{I.51}$$

2. The system of equations (I.47) - (I.51), in turn, consists of a system of 6 ordinary nonlinear differential equations (I.47) - (I.50) and one integral condition (I.51), which is the first integral of the system (I.47) – (I.50) and the initial value of the Hubble parameter.

2. We will solve the system of equations (I.47) - (I.50) with the initial conditions

$$\xi(0) = 0, \ H(0) = H_0, \ \Phi(0) = \Phi_0, \ Z(0) = Z_0, \ \varphi(0) = \varphi_0, \ z(0) = z_0. \tag{136}$$

using the circumstance that, due to the autonomy of the system (I.47) - (I.50), we can arbitrarily choose the moment of time $t_0$ at which the scale factor becomes unity $a(t_0) = 1$. In addition, the value in (136) will be determined by one of the solutions (I.51) given other initial conditions. As a result, we will set the initial conditions with an ordered list of four elements

$$\mathbf{I} = [\Phi_0, Z_0, \varphi_0, z_0]. \tag{137}$$

For simplicity, we assume $Z_0 = z_0 = 0$.

3. Due to the significant nonlinearity of the system of differential equations (I.47) - (I.50), we will find its solutions by numerical methods, using mainly the Rosenbrock method with absolute and relative accuracy of $10^{-8}$, which is applicable, among other things, to rigid systems of differential equations.

4. The system of background equations (I.47) - (I.51), which determines the unperturbed cosmological model $M_1$, is completely determined by 9 parameters, which we will specify as an ordered list:

$$\mathbf{P} = [[\alpha, m, e_z, \pi_z], [\beta, \mathfrak{m}, e_\zeta, \pi_\zeta], \Lambda], \tag{138}$$

where α, β are self-action constants; $m, \mathfrak{m}$ are masses of quanta; $e_z, e_\zeta$ are scalar charges; $\pi_z, \pi_\zeta$ are initial Fermi momentes of classical and phantom fields, respectively, Λ is the cosmological constant.

5. Thus, the coefficients $\gamma_{\alpha\beta}$ of matrix $\mathbf{A_0}$ are determined by numerical solutions of the system of background equations (I.47) - (I.51) with parameters $\mathbf{P}$ (Eq. (138)) and initial conditions (137):

$$\mathbf{A}_0 = \mathbf{A}_0(n, \mathbf{P}, \mathbf{I}, t), \ \gamma_{\alpha\beta} = \gamma_{\alpha\beta}(n, \mathbf{P}, \mathbf{I}, t). \tag{139}$$

Thus, the problem of calculating the amplitudes of perturbations $\mathbf{F}_n(z,t)$ (Eq. (120)) and their time evolution is a very complex computational problem.

### 6.2. Example of Numerical Modeling
### 6.2.1. Background solution

Let us consider the model with parameters

$$\mathbf{P}_1 = [[1,1,0.0001,0.01],[1,1,0.001,0.001],0.00001]. \qquad (140)$$

In this case, the vacuum cosmological model has singular points $M(\Phi_c, \varphi_c, H_c)$ with coordinates [4]:

$$M_{11}^{+} = [1, 1, 0.4082523729],\ M_{01}^{+} = [0, 1, 0.2886809080],\ M_{10}^{+} = [1, 0, 0.2886809080]. \qquad (141)$$

Among these points, two points $M_{11}^{+}$ and $M_{10}^{+}$ are saddle values and one $M_{01}^{+}$ is attractive. In Figs 1 and 2, one can see the results of numerical integration of the system of equations (I.47)–(I.51). The cosmological system starting at the point $M_{10}^{+}$ and passing across the point $M_{11}^{+}$, appears at the point $M_{01}^{+}$ of the stable equilibrium corresponding to inflation.

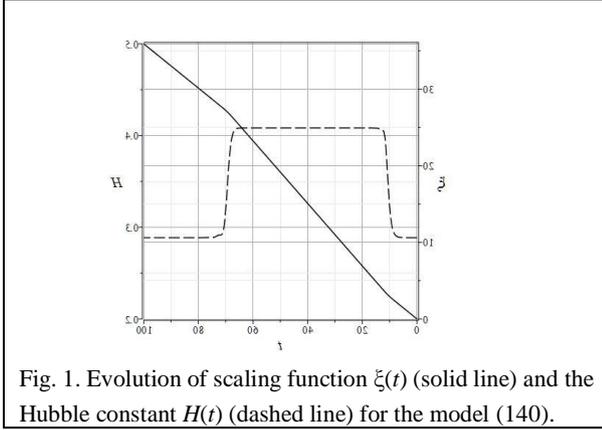
Fig. 1. Evolution of scaling function ξ(t) (solid line) and the Hubble constant H(t) (dashed line) for the model (140).

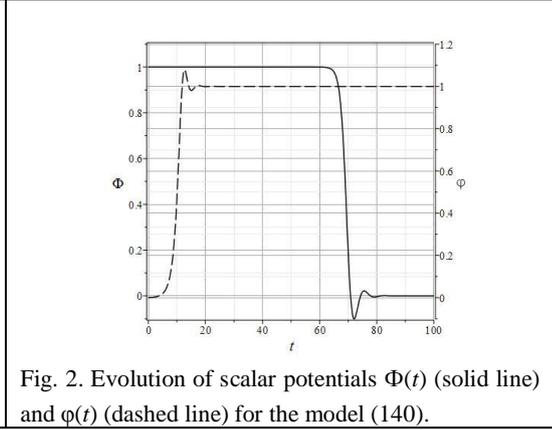
Fig. 2. Evolution of scalar potentials Φ(t) (solid line) and φ(t) (dashed line) for the model (140).

### 6.2.2. Mode 1

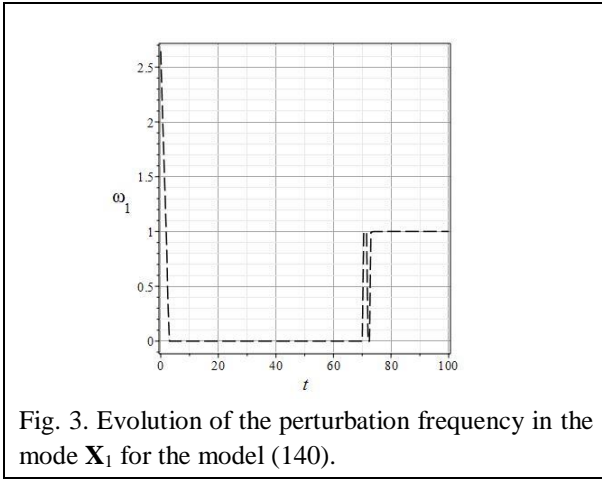
Fig. 3. Evolution of the perturbation frequency in the mode $\mathbf{X}_1$ for the model (140).

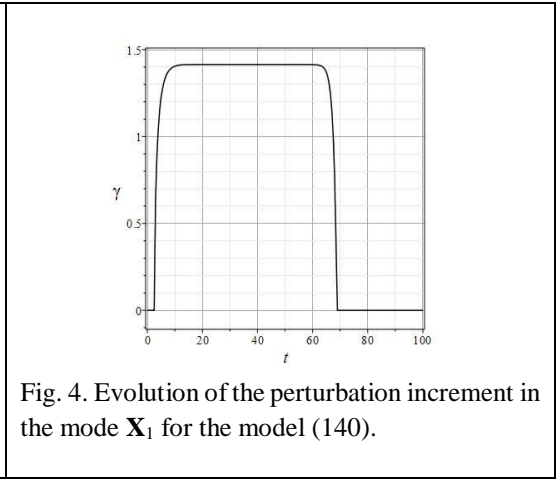
Fig. 4. Evolution of the perturbation increment in the mode $\mathbf{X}_1$ for the model (140).

Figures 3 and 4 present the evolution of the perturbation frequency and increment in the mode $\mathbf{X}_1$, and Fig. 5 shows the evolution of the perturbation components obtained by the above indicated methods[5].

---

[5] Perturbation components $\tilde{\Phi}, \tilde{\varphi}, \tilde{\nu}$ are denoted in figures as $\Phi, \varphi, \nu$.

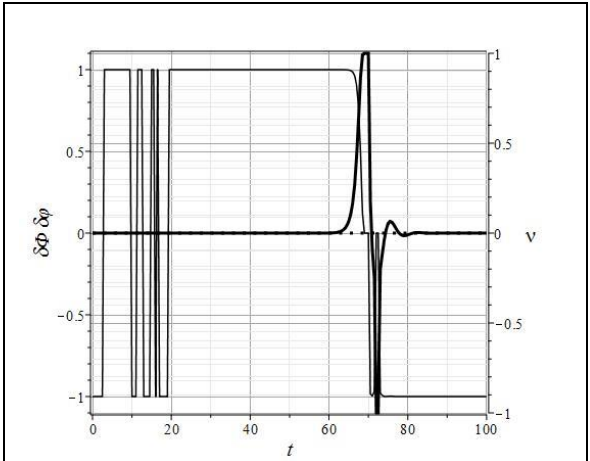

Fig. 5. Evolution of the perturbation components in the mode $\mathbf{X}_1$ for the model (140). Solid, dashed, and thick solid lines indicate $\delta\tilde{\Phi}(t)$, $\delta\tilde{\varphi}(t)$, and $\tilde{v}(t)$, respectively.

### 6.2.3. Mode 2

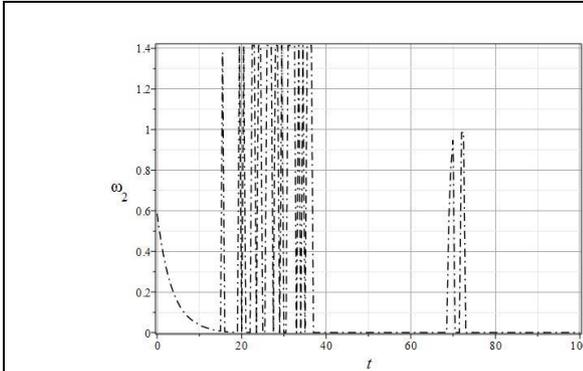

Fig. 6. Evolution of the perturbation frequency in the mode $\mathbf{X}_2$ for the model (140).

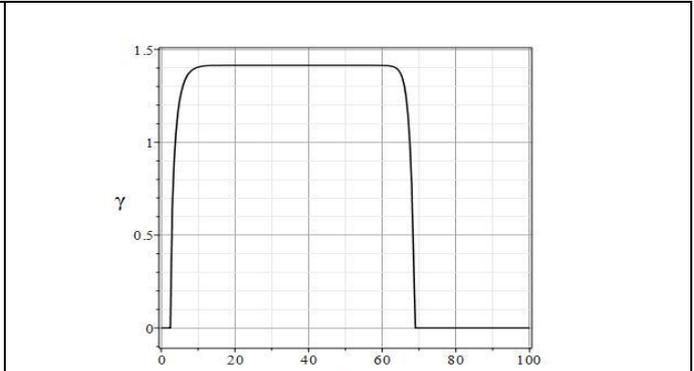

Fig. 7. Evolution of the perturbation increment in the mode $\mathbf{X}_2$ for the model (140).

Figures 6 and 7 present the evolution of the perturbation frequency and increment in the mode $\mathbf{X}_2$, and Fig. 8 shows the evolution of the gravitation component of perturbations obtained by the above indicated methods.

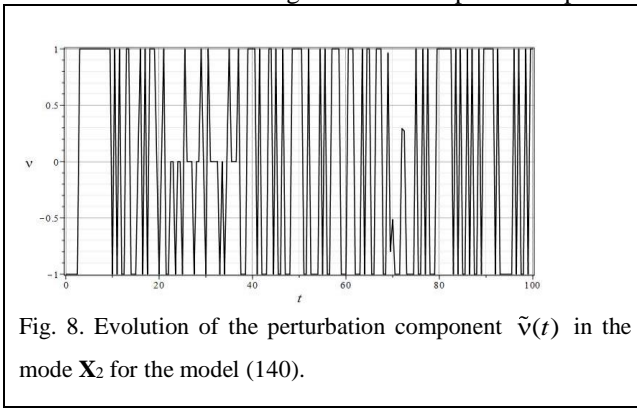

Fig. 8. Evolution of the perturbation component $\tilde{v}(t)$ in the mode $\mathbf{X}_2$ for the model (140).

### 6.2.4. Mode 3

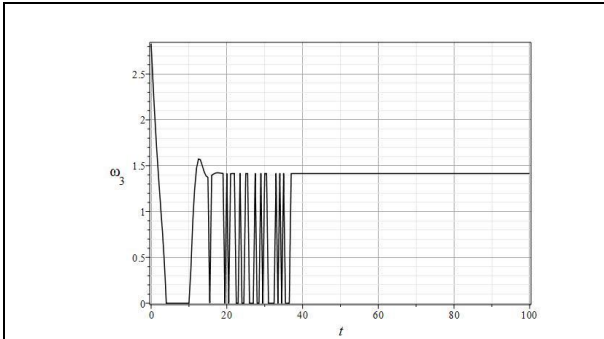

Fig. 9. Evolution of the perturbation frequency in the mode $\mathbf{X}_3$ for the model (140).

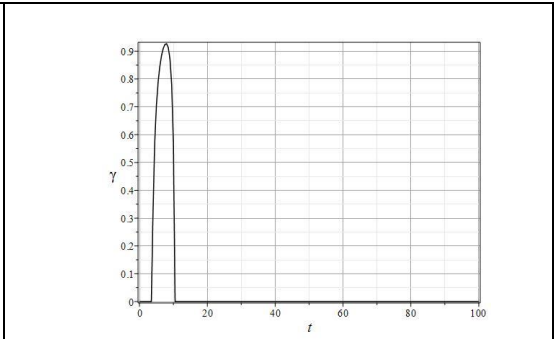

Fig. 10. Evolution of the perturbation increment in the mode $\mathbf{X}_3$ for the model (140).

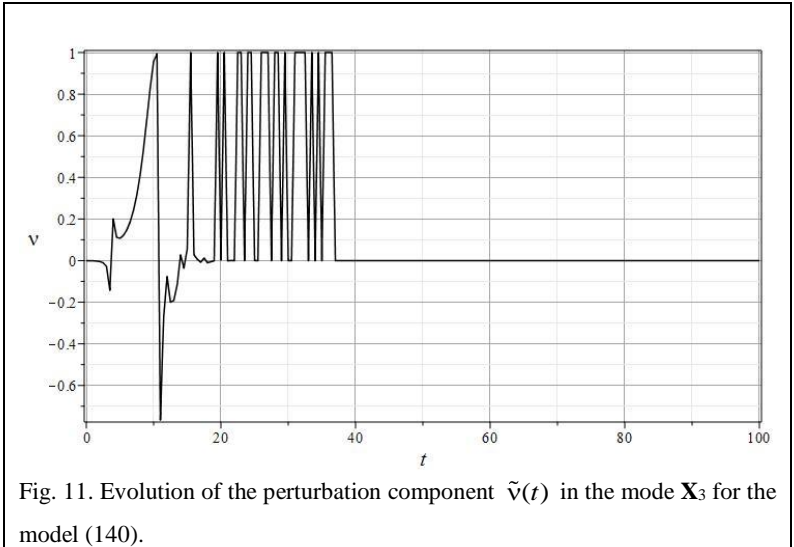

Fig. 11. Evolution of the perturbation component $\tilde{\nu}(t)$ in the mode $\mathbf{X}_3$ for the model (140).

Figures 9 and 10 present the evolution of the perturbation frequency and increment in the mode $\mathbf{X}_3$, and Fig. 11 shows the evolution of the gravitation component of perturbations obtained by the above indicated methods.

Comparing the three perturbation modes, we note, first, that the long stage of strong instability occurs only in the mode $\mathbf{X}_1$. Secondly, it can be seen by comparing Fig. 3 and 4 that at the stage of instability, the frequency of oscillations is equal to zero, which corresponds to standing growing oscillations. From the graph in Fig. 5, it can be seen that the indicated stage of instability corresponds to the growth of the potential of the classical scalar field, while the phantom field is practically absent. We also note the aperiodic nature of the oscillations of the gravitational field in the mode $\mathbf{X}_2$ (Fig. 8). In the mode $\mathbf{X}_3$, the frequency of oscillations after aperiodic oscillations becomes constant (Fig. 9) precisely at the moment when gravitational perturbations disappear (Fig. 11).

## 7. RESULTS AND DISCUSSION

Summing up the results of the study, we note its most important results.

1. A mathematical model is constructed for the evolution of perturbations in a cosmological model based on a two-component statistical system of scalarly charged degenerate fermions interacting through scalar, classical, and phantom fields with Higgs potentials. This model consists of a system of nonlinear background equations of a homogeneous cosmological model and a system of linear homogeneous differential equations describing the time evolution of perturbations against the background of a homogeneous isotropic cosmological model (evolution equations). Based on evolutionary equations, the closed system of homogeneous algebraic equations was created in the WKB approximation, that described the evolution of short-wave perturbations. The system coefficients were determined by the functions of background solutions.

2. In the WKB approximation, on the basis of evolution equations, a closed system of homogeneous algebraic equations is constructed that describes the evolution of short-wave disturbances. The coefficients of the system of equations are determined by the functions of the background solutions.

3. The solution of the system of these equations is reduced to the problem of finding eigenvectors and eigenvalues of the matrix of a linear operator. Relationships between perturbation functions and eigenvectors and values of the system matrix are found.

4. A program for the numerical solution of the problem of the evolution of perturbations was created in the applied mathematical package. The application of the program made it possible to separate the contributions in three perturbation modes.

5. An example of numerical simulation of the evolution of perturbations is given and analyzed.

Note that in this part of the article we have only given an example of numerical simulation of the evolution of perturbations, which demonstrates the gravitational-scalar instability. In the next part of the article, we present the results of modeling a wide class of cosmological models and analyze them for their suitability for explaining the nature of the formation of black holes in the early Universe.

**Funding**

This paper has been supported by the Kazan Federal University Strategic Academic Leadership Program.